\documentstyle[jkas]{article}

\beginpage{1}
\endpage{25}
\year{2011}\volume{xx}\month{???}

\runningauthor {KIM ET AL.}
\runningtitle{Auto-Guiding System for CQUEAN}

\month{December} \year{2010} \volume{43} \issueno{5}
\beginpage{1}\endpage{25}
\date{Received January 10, 2011; Accepted XXX, 2011}

\begin{document}
\title{Auto-Guiding System for CQUEAN (Camera for QUasars in EArly uNiverse)}
\author{Eunbin Kim$^1$, Won-Kee Park$^2$, Hyeonju Jeong$^1$, Jinyoung Kim$^1$, John Kuehne$^3$, Dong Han Kim$^4$,\\
   Han Geun Kim$^4$, Peter S. Odoms$^3$, Seunghyuk Chang$^5$, Myungshin Im$^2$, and Soojong Pak$^{1}$}
\address{$^1$ School of Space Research, Kyung Hee University,
1 Seocheon-dong, Giheung-gu, Yongin-si, Gyeonggi-do 446-701, Korea\\
{\it E-mail : ebkim@khu.ac.kr, jhyeonju@khu.ac.kr, jinyoungkim@khu.ac.kr, and soojong@khu.ac.kr}}
\address{$^2$ CEOU/Dept. of Physics and Astronomy, Seoul National University, Seoul 151-742, Korea\\
{\it E-mail : wkpark@astro.snu.ac.kr and mim@astro.snu.ac.kr}}
\address{$^3$ McDonald Observatory, Univ. of Texas at Austin, Austin, TX, 78712-0259, USA\\
 {\it E-mail : jwkuehne@astro.as.utexas.edu and pso@astro.as.utexas.edu}}
\address{$^4$ Department of Electrical Engineering, Kyung Hee University, Gyeonggi-Do 446-601, Korea\\
{\it E-mail : donghani@khu.ac.kr and sskhk05@khu.ac.kr}}
\address{$^5$ Samsung Electronics, Suwon, Gyeonggi-do 443-370, Korea\\
{\it E-mail : chang@offaxis.co.kr}}
\address{\normalsize{\it (Received January 10, 2011; Accepted XXX X, 2011)}}
\offprints{Won-Kee Park}

\abstract{
To perform imaging observations of optically red objects such as high redshift quasars and brown dwarfs,
the Center for the Exploration of the Origin of the Universe (CEOU) recently developed an optical CCD camera, Camera for QUasars in EArly uNiverse (CQUEAN), which is sensitive at 0.7-1.1 $\mu$m. To enable observations with long exposures, we developed an auto-guiding system for CQUEAN. This system consists of an off-axis mirror, a baffle, a CCD camera, a motor and a differential decelerator. To increase the number of available guiding stars, we designed a rotating mechanism for the off-axis guiding camera. The guiding field can be scanned along the 10 arcmin ring offset from the optical axis of the telescope. Combined with the auto-guiding software of the McDonald Observatory, we confirmed that a stable image can be obtained with an exposure time as long as 1200 seconds.}

\keywords{instrumentation: CCD camera --- techniques: photometric --- telescopes --- auto-guiding}
\maketitle

\section{INTRODUCTION}

Most telescopes require active guiding to stay on target for long periods. Atmospheric refraction, mechanical imperfections, flexure of the mount, and other errors make long unguided exposures difficult  \citep{lee07}. Historically, astrophotographers guided by watching a star through an eyepiece while manually adjusting the position of the telescope or camera. Today guide cameras with charge-coupled device (CCD) sensors and computer control have revolutionized astronomy, automatically guiding  for long durations and with better accuracy than is humanly possible \citep{bir06}.

In the auto-guiding system, an image of the guide star is taken repeatedly, and a computer calculates the centroids and commands the telescope control system (TCS) to compensate for drift.  On some instruments the guide star is the actual science target - for example the image on a spectrograph slit - but our guider captures the image from outside the science field with a separate movable guide camera. Another design uses a separate telescope for the guider.

In Korea, auto-guiding systems have been developed to improve tracking accuracy \citep{jeo99, yoo06}, and \citet{mun06} developed a telescope control program which has auto-guiding capability for the telescope in Kyung Hee University. Efforts to improve guiding accuracy \citep{ise08} and pointing accuracy \citep{kan06} have been made at other observatory. 
Commercial auto-guiding programs are even popular among amateur astronomers.

\begin{figure}[t]
\plotfiddle{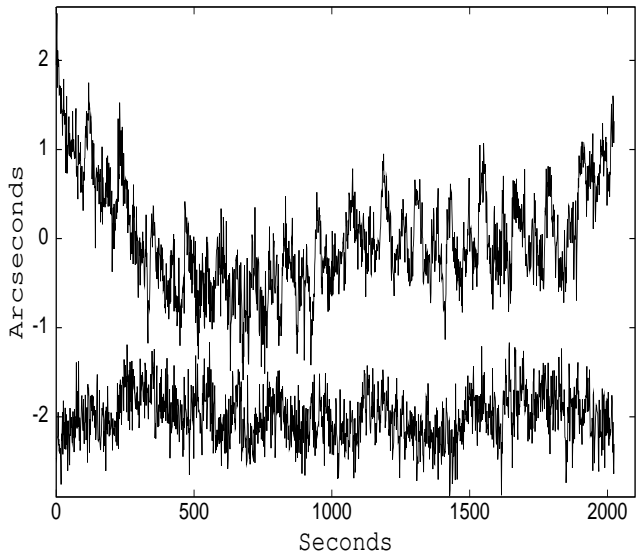}{8cm}{270}{34}{44}{-140}{260}
\caption{Unguided drift of the 2.1m telescope along Right Ascension (top) and Declination (bottom) direction. It shows the 2-minute periodic error in RA caused by a slight misalignment of the final worm gear. The longer period motion is probably caused by slight imperfections in the gear teeth and flexure in the mount. High frequency motion in Dec may be caused by wind shake and oscillations of the mount that disturb the image centroid, and presumably the motion in RA contains the same kind of noise. The source of longer period motion in Dec is unknown.}
\label{drift}
\end{figure}

CQUEAN is an optical camera system for the survey of high redshift ($z>5$) quasars in the early universe, which is developed by the Center for the Exploration of the Origin of the Universe (CEOU). It is composed of a science CCD camera, a focal reducer, seven broad-band filters (g', r', I', z', Y, Iz and Is) and an auto-guiding system. The science camera has a deep-depletion CCD whose sensitivity at 1um is better than conventional designs. It is attached to the 2.1m Otto Struve Telescope at McDonald Observatory, Texas, USA. The overall description of CQUEAN will be presented in \citet{par11}.

In the high and dry desert of far West Texas, the McDonald Observatory is an excellent location for astronomical observations. The 2.1m Otto Struve Telescope is a classical cassegrain telescope built in 1939 by Warner \& Swasey, and was the second largest telescope in the world. The cassegrain focal ratio is f/13.65. Like most telescopes of its era, it rotates on a equatorial mount.

An auto-guider was developed by \citet{abb90}, but it was not successful. At present, one of the main instruments on the 2.1m telescope, the Sandiford Echelle Spectrograph, is now fully auto-guided, but the hardware and software is quite specific to the spectrograph. Thus we developed an auto-guiding system for CQUEAN in order to observe faint targets with long exposures.

In this paper, we describe the development of the CQUEAN auto-guiding system and analyze its guiding performance. In section 2, we show the tracking status of the 2.1m Otto Struve Telescope without auto-guiding, the sensitivity of the guide camera, and the expected number of guide stars for the estimated sensitivity. In section 3, the system design and the components of the auto-guiding system are described. In section 4, we present the analysis of the first observations, and make some concluding remarks in section 5.

\section{SYSTEM REQUIREMENTS}

\subsection{Tracking Errors}

Tracking is accomplished through a cascade of three worm gears, originally designed to rotate the polar axis at
$15.037 \arcsec$/sec when driven by an 1800 rpm motor. Today the final worm is precisely controlled by a closed-loop servo system, and the telescope track rate is fully adjustable.
However, the polar axis is not part of the closed-loop system, and its motion depends on the minute interactions of the final worm with its 720-tooth bronze gear.

Fig. \ref{drift} shows the telescope drift along the Right Ascension (RA) and Declination (Dec) axes. Fourier analysis reveals a periodic error with an amplitude of about 1 arcsecond and period of about 2 minutes.
We now know that a slight misalignment of the worm relative to its gear causes the telescope to nod west when a new tooth enters the worm, and to the east when a tooth leaves the worm. Thus the motion looks like a triangle-wave.

About half of the periodic error can be removed by simply subtracting the motion through modeling or guiding corrections made every few seconds. There are longer-period variations caused by imperfections in the 720-tooth gear that are easily removed by guiding. In certain conditions we also see small high frequency slip-stick motions that are probably caused by bearing friction, but these cannot be removed by guiding. For most observations guiding in declination is not important, and only requires a slow drift of about $0.001 \arcsec$/sec. Tracking tests from 2009 show the best performance is achieved when corrections are made every few seconds.

\subsection{Number of expected guide stars and guide camera field of view}

\begin{figure}[t]
\centering \epsfxsize=8cm
\epsfbox{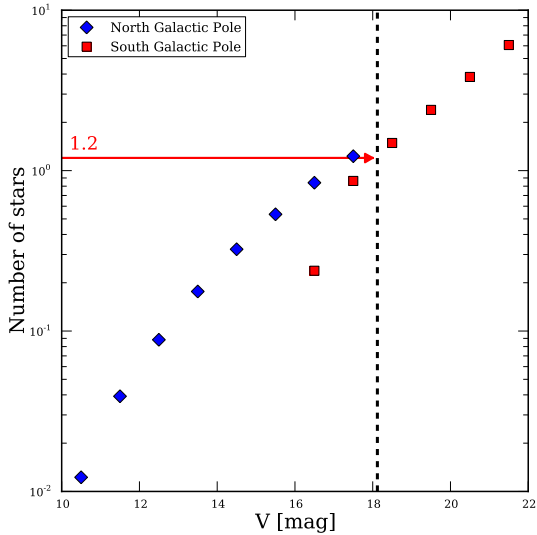}
\caption{Expected number of guiding stars in the guide CCD field ($2.97\arcmin \times 2.97 \arcmin$) that can be detected with $10\sigma$ limit in 1 sec exposure. The blue diamonds in the graph represent the number of stars in the north galactic pole and the red squares the stars in the south galactic pole. Vertical dashed line indicates the limiting magnitude for our guide CCD camera.}
\label{starcount}
\end{figure}

\begin{figure}[!t]
\centering
\epsfxsize=7cm
\epsfbox{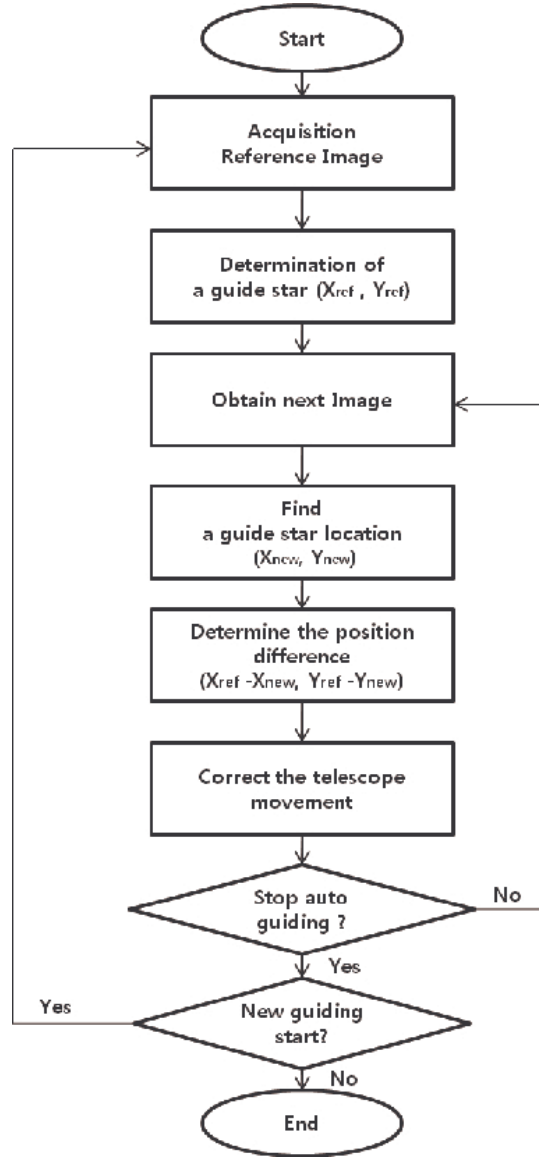}
\caption{Auto-guiding procedure of CQUEAN.}
\label{procedure}
\end{figure}

To obtain corrections every few seconds requires a fairly bright guide star. We estimated the expected number of guide stars to check the feasibility of the auto-guiding system. Since the guide camera must take short exposures of bright stars, system noise affects the sensitivity more than sky background. Thus we assume that the sensitivity of the guide camera is limited by the system noise, and calculate the detection limit with S/N of 10 with 1 sec exposure time. The detection limit under high system noise (McLean 2008) is given as

\begin{equation}
m = m_{zp} - 2.5log \left( {1 \over g} \cdot {S\over N} \cdot {R\over T} \cdot \sqrt{ {N_{pix} \over n_{0}}} \right)
\end{equation}
where $m_{zp}$ is the zeropoint, $g$ the gain ([electrons/ADU]), and $S/N$ represents signal to noise ratio. $N_{pix}$ is the number of pixels that cover a stellar profile and $n_0$ is the number of exposures. T is the exposure time in seconds and R is the readout noise in electrons.
The zeropoint of the system is calculated to be 23.18 mag with the assumptions that total system throughput is 0.8 and CCD gain is 10. For the readout noise, we use the value of 8 electron/pixel from the camera manufacturer's test report. We assume $N_{pix}$ to be 149.42 pixels based on the f-ratio of the telescope and median value of $1.2\arcsec$ for full width at half maximum (FWHM) seeing in V at McDonald Observatory.

With these assumptions, we obtain $m_v = 18.17$ for the operational requirement of the guide camera. Using the star count data at galactic poles \citep{all00}, we calculated the expected number of stars brighter than the limiting magnitude over a $2.97\arcmin \times 2.97\arcmin$ field of view. Fig. \ref{starcount} shows that about 1.2 stars may fall on our guide camera field of view. The number will be higher than this value at the lower galactic regions. There will be nights with bad seeing and clouds, so we require that there should be at least five stars suitable for guiding ($S/N > 10$) in the guiding field. To satisfy this requirement, we designed a moving mechanism to rotate the guide camera around the optical axis, which increases the guiding field of view by about a factor of five.
This is more advantageous than adopting a large format CCD at a fixed position, since aberration will be introduced if the guide field of view is too large.

\section{CQUEAN AUTO-GUIDING SYSTEM}
With the requirements for auto-guiding at hand, we designed our system. Fig. \ref{procedure} shows the auto-guiding procedure.

\begin{figure}[!t]
\plotfiddle{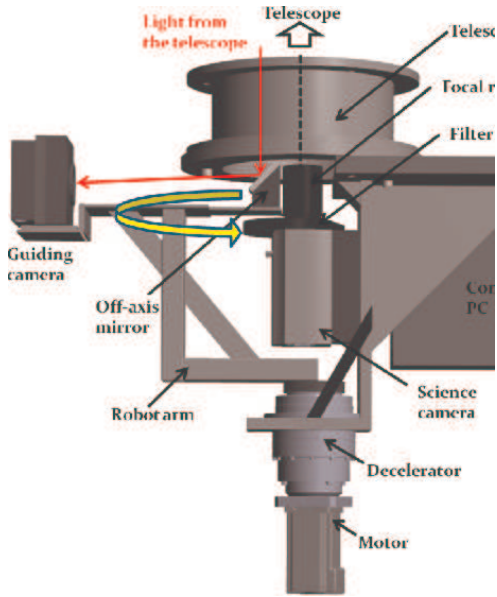}{8cm}{270}{40}{40}{-110}{230}
\caption{3D Design of CQUEAN structure (without the baffle and the cover).}
\label{3D}
\end{figure}

\subsection{Overview of the guiding system}

The overview of the opto-mechanical part in the auto-guiding system is given in Fig. \ref{3D}. The telescope adapter holds all components of CQUEAN: science camera, guide camera, filter wheel, and a control PC. The incoming light of the guide field is sent from the telescope secondary to the off-axis mirror and it is reflected into the guide camera. If no suitable guide star is found, the guide camera and the off-axis mirror can be rotated with the arm around the telescope's optical axis. The auto-guiding software acquires an image every 1-5 sec, but a shorter period is possible (see also Fig. \ref{guiding}, and the contents in Section 4.2.2). In the following subsections, we describe key hardware components and software components of the auto-guider.

\begin{table}[t]
\begin{center}
\centering
\caption{Specification of FLI PL1001E on the 2.1m telescope \label{tbl1}}
\doublerulesep2.0pt
\renewcommand\arraystretch{1.5}
\begin{tabular}{cc}
\hline \hline
Pixel size      & 24$\mu$m$\times$24$\mu$m   \\
Pixel format    & 1024$\times$1024   \\
Pixel scale     & 0.174$\arcsec$/pixel \\
Field of view   & 2.97$\arcmin\times$2.97$\arcmin$ \\
CCD type        & Front illuminated \\
Readout noise   & 8.70e-RMS @ 1MHz\\
Bias            & 1850 counts \\
Peak QE         & 72\% \\
Cooling method  & Air cooling \\
Dark current    & $\le$ 0.2e-/pixel/sec @ -45$\arcdeg$C\\
Linear full well& 500,000 e-\\
Readout speed & 1 MHz, 3.3 MHz \\
Gain          & 2.42 electrons/ADU \\
\hline
\end{tabular}
\end{center}
\end{table}

\subsection{Guide camera}

A ProLine1001E CCD camera from Finger Lakes Instrumentation Co. is used for the guide camera. This camera employs a front illuminated KODAK KAF-1001E detector, which has $1024 \times 1024$ pixels of $24\mu m\times 24 \mu m$ size. The pixel scale is $0.174\arcsec$/pixel, and the field of view is $2.97\arcmin \times 2.97\arcmin$ at the cassegrain focus of the 2.1m telescope. Table 1 summarizes the basic specifications of our guide camera.
The camera is cooled to $-30\arcdeg$C during normal operation via air cooling. The camera housing includes the CCD detector as well as the cooling unit and supporting electronics. It is connected to the control computer via Universal Serial Bus (USB), and the readout speed is 1 MHz, which is more than enough for our auto-guiding application.

\subsection{Guide camera system platform}


\begin{figure}[!t]
\centering \epsfxsize=7cm
\epsfbox{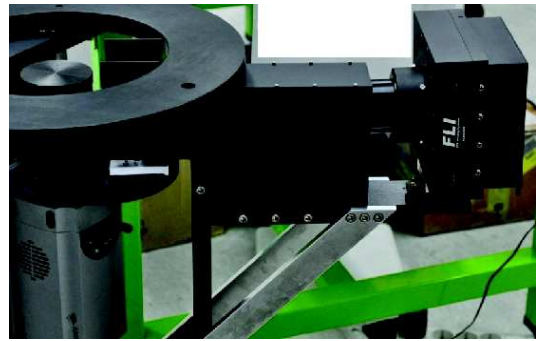}
\caption{CQUEAN auto-guiding camera system. The black box in the center is the cover
for the guide CCD camera, which encloses the camera baffle at right end. Note off-axis
mirror is blocked from the view.}
\label{system}
\end{figure}

As described in the previous section, the guide camera system has to be movable to see different off-axis fields. Therefore, the guide camera and the off-axis mirror are attached to an arm that rotates around the center of the optical axis. The guide camera is installed about 300mm away from the optical axis, and it is 2.12 degrees tilted forward to make the image plane vertical to the central beam. The position of the guide camera can be adjusted for focus. A baffle and cover are installed to block stray light from entering the guide camera.
Fig. \ref{system} shows the rotating arm and the guide camera system on it. The rotating arm consists of three main struts and two supporting braces to minimize flexure. The motor system consists of an MDRIVE 34 motor from IMS and the differential decelerator, AD140-050-P1 from APEX, and its commands are sent over a USB-serial converter.
On the 2.1m telescope, the center of the guiding field is separated from the center of the science field of view by $10\arcmin$. The arm is allowed to move azimuthally from $-20\arcdeg$ to $70\arcdeg$ which is the range that avoids mechanical interference with other parts of the CQUEAN system. Therefore, the total available field of view is about 63 square arcminutes, satisfying the requirement for the number of available guide stars.

\subsection{Software}

\begin{figure}[!t]
\plotfiddle{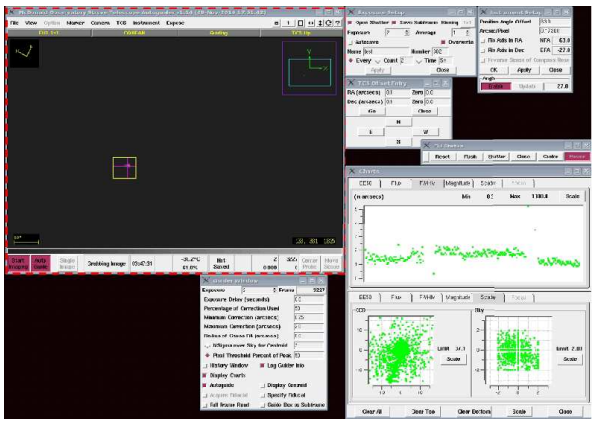}{6.5cm}{270}{28}{28}{-120}{170}
\caption{Display screen of $agdr$, the guiding control program. Red dashed box denotes main window.}
\label{agdr}
\end{figure}

\begin{figure}[t]
\plotfiddle{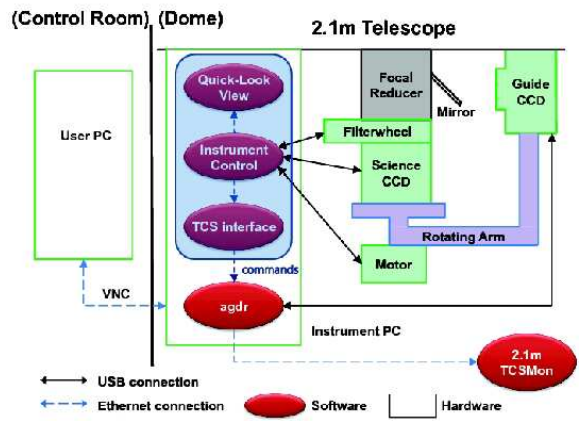}{6.5cm}{0}{40}{40}{-120}{0}
\caption{Block diagram of all CQUEAN components.}
\label{diagram}
\end{figure}

Two programs are used for auto-guiding CQUEAN. The arm motion is controlled by the \emph{Data Taking Package}, which is the main control program of CQUEAN. The auto-guider, $agdr$, was developed at the McDonald Observatory and was specially modified for CQUEAN's guide camera, including the angle of the arm supplied by the \emph{Data Taking Package}. In addition to auto-guiding, it continuously monitors the seeing condition of the guide star.  Fig. \ref{agdr} displays several of \emph{agdr}'s windows.
 The interactions between the two programs, as well as all hardware components of CQUEAN, are shown with block diagrams in Fig. \ref{diagram}.


\section{THE OBSERVATION AND ANALYSIS}

\subsection{Observation Log}

CQUEAN was commissioned and tested August 10-17, 2010 on the 2.1m telescope. Various targets such as standard stars, globular clusters, quasars, and gamma-ray bursts were observed with the science camera to quantify its performance. Skies were clear except for one night and Table 2 lists the seeing conditions. We varied the exposure times from 0.1 to 1200 seconds to test the performance of the auto-guiding system and the science camera.  Additional tests were carried out during the subsequent run on December 20-29, 2010.

\begin{table}[t]
\scriptsize{\tiny}
\begin{center}
\centering
\caption{Weather condition during the observation\label{tbl2}}
\doublerulesep2.0pt
\renewcommand\arraystretch{1.5}
\begin{tabular}{cccc}
\hline\hline
         &\multicolumn{2}{c}{FWHM of PSF$^{\rm a,b}$}& \\
\cline{2-3}
Date(UT) & Science camera & Guiding & Sky\\
          & i'-filter & camera& \\
\hline
2010 Aug 12 &  -   & 1.24 & - \\
2010 Aug 13 & 0.96 & 1.22 & - \\
2010 Aug 14 & 1.02 & 1.35 & - \\
2010 Aug 15 & 1.07 & 1.44 & thin cirrus \\
2010 Aug 16 & 1.07 & 1.23 & cloudy \\
2010 Aug 17 & 0.83 & 1.06 & clear \\
2010 Aug 18 & 0.85 & 1.03 & very clear \\
\hline
\end{tabular}
\end{center}
\begin{tabnote}
\hskip18pt $^{\rm a}$Units in arcsec.\\
\end{tabnote}
\begin{tabnote}
\hskip18pt $^{\rm b}$Average value during the night.
\end{tabnote}
\end{table}


\begin{figure}[h]
\centering \epsfxsize=8cm
\epsfbox{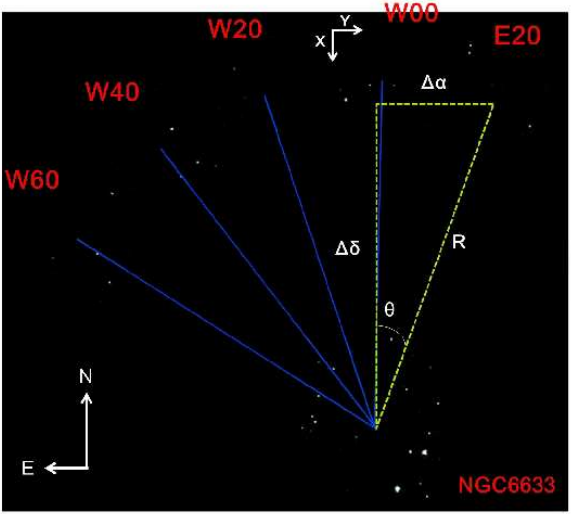}
\caption{Locations of science CCD camera field and the guiding CCD camera fields. 
The XY arrows on W00 image indicate the actual orientation of guide CCD chip. Note the guide field names are after the locations of guide CCD camera, not the locations of fields. See section 4.2 for explanation.}
\label{location}
\end{figure}

\subsection{Operation Test}

\begin{deluxetable}{cccccc}
\tablecolumns{6}
\tablewidth{0pc}
\tablecaption{Guiding image information\tablenotemark{\dagger}\label{tbl3}}
\tablehead{
\colhead{Field} &
\colhead{Angle} &
\colhead{RA} &
\colhead{Dec} &
\colhead{RA axis angle\tablenotemark{a}}&
\colhead{Dec axis angle\tablenotemark{b}}
}
\startdata
E20 &-20 & 276.86590 & 6.72022 & 68.894 & 248.836 \\
W00 & 0  & 276.92226 & 6.73242 & 88.970 & 268.943 \\
W20 & 20 & 276.97954 & 6.72482 & 108.801 & 288.867 \\
W40 & 40 & 277.03124 & 6.69820 & 128.915 & 308.871 \\
W60 & 60 & 277.06961 & 6.65577 & 148.927 & 328.924 \\
\enddata
\tablenotetext{\dagger}{All values are in unit of degree.}
\tablenotetext{a}{The angle of east axis with respect to X axis of the chip in clockwise direction}
\tablenotetext{b}{The angle of north axis with respect to Y axis of the chip in clockwise direction}
\end{deluxetable}

To measure the exact location of the guide camera observation field with respect to the science camera we observed a region in NGC 6633 on August 17, 2010. Different guiding fields were observed with the science CCD camera fixed to a position by rotating the arm by $20\arcdeg$ in each step in order to check the performance of the moving mechanism. It turned out that the guide camera sees an off-axis field in opposite direction with respect to the optical axis of the telescope, i.e., it sees northwest off-axis field when the camera is at southeast off-axis position. Fig. \ref{location} shows the locations of several guiding images with respect to the scientific field, and Table 3 lists the coordinates of the guiding field. Each field is named after the location of guide camera. The science image center is RA=276.9260$\arcdeg$ and DEC=6.56747$\arcdeg$. The separation between the science image and the guide-field images is about 10$\arcmin$.

\subsubsection{Hardware Performance}

Using the celestial coordinates of each image center, we calculate R, the distance between the science image center and a guide image center, and $\theta$, the angle between successive positions of the rotating arm, as indicated in Fig. \ref{location}. Eqs.(2) \& (3) derived from spherical trigonometry, were used to calculate both values. 


\begin{equation}
 R^2 \approx ({\delta_g-\delta_s})^{2} + {(\alpha_g-\alpha_s)\cos{(\frac{\delta_g+\delta_s}{2}})}^{2}\\
\label{Req}
\end{equation}

\begin{equation}
 \cos\theta = \frac{\cos{\Delta\alpha}-\cos{\Delta\delta}\cos{R}}{\sin{\Delta\delta}\sin{R}}\\
\label{Teq}
\end{equation}

where $(\alpha_{s}, \delta_{s})$ and $(\alpha_{g}, \delta_{g})$ denote the RA and Dec of science CCD image center and guide CCD image center, respectively. $\Delta\alpha$ and $\Delta\delta$ are defined as:

\begin{equation}
 \Delta\alpha = (\alpha_g-\alpha_s)\cos{\delta_g}\\
\end{equation}
\begin{equation}
 \Delta\delta = \delta_g -\delta_s\\
\end{equation}

\begin{deluxetable}{cccccccc}
\tablecolumns{8}
\tablewidth{0pc}
\tablecaption{Angles and separations of the images\label{tbl4}}
\tablehead{
\colhead{}&
\colhead{Ideal}&
\multicolumn{3}{c}{Original data}&
\multicolumn{3}{c}{Re-centered data\tablenotemark{a}}\\
\cline{3-8}\\
\colhead{Frame}&
\colhead{Rotation}&
\colhead{Rotation}&
\colhead{$\theta-\Phi$}&
\colhead{Image}&
\colhead{Rotation}&
\colhead{$\theta^{\prime}-\Phi$}&
\colhead{Image}\\
\colhead{}&
\colhead{angle($\Phi$)\tablenotemark{b}}&
\colhead{angle($\theta$)\tablenotemark{b}}&
\colhead{}&
\colhead{separation(R)\tablenotemark{c}}&
\colhead{angle($\theta^{\prime}$)\tablenotemark{b}}&
\colhead{}&
\colhead{separation($\mathrm{R^{\prime}}$)\tablenotemark{c}}
}
\startdata
E20 &-20 & -21.3520 & -1.3520 & 9.841  &-22.3162 & -2.3162 & 9.907 \\
W00 & 0  &  -1.3002 & -1.3003 & 9.899  & -2.3342 & -2.3342 & 9.905 \\
W20 & 20 &  18.6613 & -1.3387 & 9.965  & 17.6831 & -2.3169 & 9.909 \\
W40 & 40 &  38.6322 & -1.3678 & 10.042 & 37.8288 & -2.1712 & 9.931 \\
W60 & 60 &  58.2289 & -1.7711 & 10.065 & 57.6880 & -2.3120 & 9.913 \\
\cline{1-8}
Standard  & - & - & 0.195 & 0.094 & - & 0.067 & 0.011 \\
deviation &   &   &       &       &   &       &    \\
\enddata
\tablenotetext{a}{After adjusting the science center by ($\Delta\alpha=0.003\arcdeg$, $\Delta\delta=0.00001\arcdeg$.)}
\tablenotetext{b}{Units in degree.}
\tablenotetext{c}{Units in arcmin.}
\end{deluxetable}

\begin{figure}[t]
\centering \epsfxsize=8cm
\epsfbox{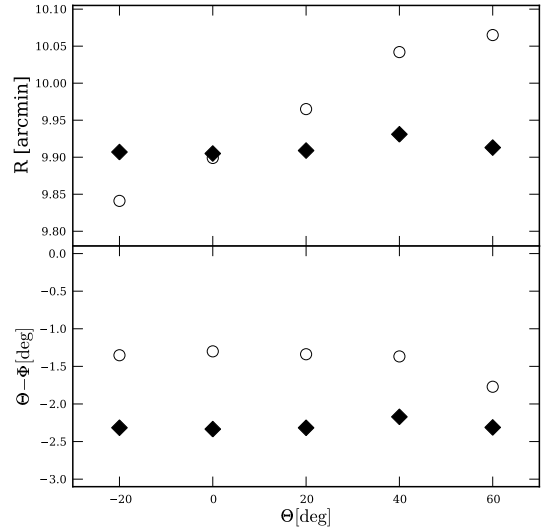}
\caption{R, the distance between the centers of science CCD field and guide
CCD field (top), and difference between the expected and measured step angles (bottom),
according to the rotation angle of the arm. Open circles denote
the values with respect to the original center of science CCD, while
filled diamonds indicate the values with respect to the new center.}
\label{anglestep}
\end{figure}

The results are listed in Table \ref{tbl4}. Of interest, R is increasing as the arm rotates further west, apparently because the center of rotation of the arm does not exactly coincide with that of the science camera. Therefore, we located the science image center that minimizes the standard deviation of R values. The new center position is located at $\Delta\alpha$ = 0.003$\arcdeg$, $\Delta\delta$ = 0.00001$\arcdeg$ apart from the original center. The R$\arcmin$ and $\theta\arcmin$ represent the values of R and $\theta$ on the re-centered data, and listed in the columns 6 to 8.
Fig. \ref{anglestep} shows $\theta-\Phi$ and R with respect to each center.

As for $\theta$, we note there is about -1.3$\arcdeg$ offset between the expected and the measured values, indicating the reference position of the guide camera frame, W00, is not exactly aligned with the meridian of the science image. We also see there are variations of about 0.195$\arcdeg$ and 0.067$\arcdeg$ for the old and new centers respectively. The step size for the combined system of the motor and the decelerator can be calculated from the specifications of each device, i.e. ${\Delta\theta}_{step}$ = 0.0035$\arcdeg$. Also the backlash of the decelerator is $5\arcmin(\approx 0.0083\arcdeg$). Considering these factors, we surmise that the standard deviation of differences between re-centered and ideal rotation angle is affected by backlash of the decelerator.

In the case of R, variations are probably due to mechanical flexure of the rotating arm. The standard deviation of R at the re-centered position is 0.011$\arcmin$ (0.66$\arcsec$) for 80$\arcdeg$ rotation of the guiding system.

\begin{figure*}[t]
\plotfiddle{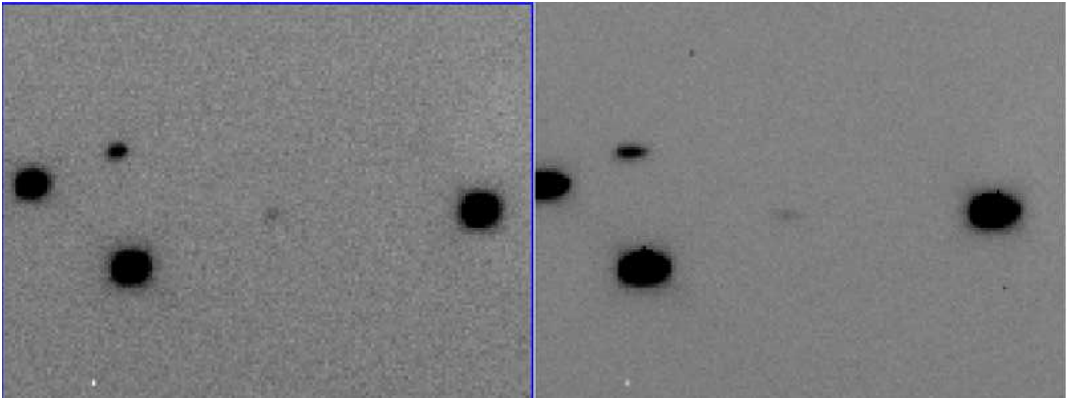}{7cm}{270}{50}{50}{-200}{200}
\caption{Comparison of 600 sec. images with auto-guiding (left) and without auto-guiding (right). The field of view for the image in each panel is $4.3\arcsec \times 4.3\arcsec$}
\label{comp}
\end{figure*}

\subsubsection{Analysis of Guiding Performance}

We examined profiles of astronomical sources in our science images to estimate guiding performance. Perfect auto-guiding would make circular stellar profiles, and would leave no trails in long exposures. To demonstrate how no guiding affects the image quality, we compare two 600 second images with auto-guiding (left) and without auto-guiding (right) in Fig. \ref{comp}. These images are taken on the same night on the same target successively. The seeing condition was identical (FWHM $\sim 0.8\arcsec$) judging from other images taken just before and after this test observation. The stellar profiles on the auto-guided image are rounder with ellipticity $\sim 0.1$, while those in non-auto-guided image are quite elongated in the RA direction with ellipticity $\sim 0.4$ as expected from the tracking error analysis as described in section 2.1. Also faint stars appear more vaguely on the non-auto-guided image, since the fluxes of stars are spread over a larger area on the CCD.

To analyze the guiding performance in a more quantitative way, we selected i-band images of various exposure times from 1 sec to 1200 sec in order to compare the PSF shapes with each other. 
SExtractor \citep{ber96} was ran on all i-band images taken during August 10-17, 2010, except for the ones obtained during focusing procedures, to obtain the FWHMs, ellipticity, and Position Angles (PA) of the various sources on them. The PA in this study is defined as the angle between the major axis of a source profile and X axis of the science camera image, and it ranges between $-90\arcdeg \sim 90\arcdeg$. The X axis of chip lies $1.1\arcdeg$ clockwise direction with respect to east RA axis for CQUEAN. We tried to include  only point sources in the analysis with conditions of stellarity $> 0.95$, and flag $=0$. All point sources with FWHM $\ge 6$ pixel were visually checked, which revealed they are either saturated or cosmic rays. \citet{par11} shows that FWHM and ellipticity show weak tendencies of increasing in regions far from the image center due to aberration. Therefore, we limit the analysis samples within 200 pixel radius from the image center. 

\begin{figure}[t]
\centering \epsfxsize=8cm
\epsfbox{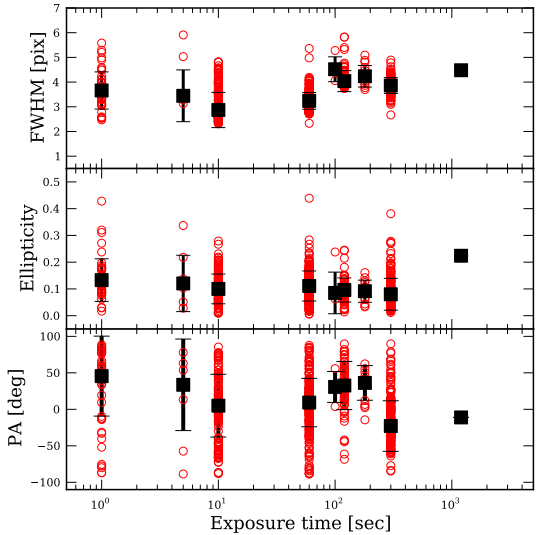}
\caption{FWHM (top), ellipticity (middle) and Position Angles (PA) (bottom) of stellar image profiles against exposure time. See section 4.2.2 for the definition of PA in this study. Red open circles denote the individual measured values, and black squares denote the median value of the data. Error bars indicate the 1-$\sigma$ level of the data.}
\label{fwhm}
\end{figure}

The FWHM, ellipticity, and PA of final samples are plotted against exposure time in Fig. \ref{fwhm}. In all panels, the black squares and error bars indicate the median and 1-$\sigma$ level of the samples, respectively.
The medians of FWHM remain around $\sim 4$ pixels, without showing any increasing tendency, up to 300 sec. exposure. 
Same tendency can be seen for the medians of ellipticity against exposure time. The fact that all medians of ellipticity fall around 0.1 indicates that most of the point source profile is quite circular even for exposures as long as 300 sec.
Meanwhile, the PA distribution shows very large scatter over whole range of PA for all exposure time data. The incorrect guiding of telescope would make stellar profile elongated along RA axis as shown in right panel of Fig. \ref{comp}, and PA distribution would show a strong peak around $1.1 \arcdeg$. However, large scatter of PA and its random median values over the range of exposure time, combined with small ellipticity, again prove that point source profiles on our images are quite circular even for long exposures. Note that 1200 sec. exposure sample shows different values in all panels. However, we could not make significant conclusion due to few available data. Based on all analyses, we conclude that our auto-guiding system works quite well.



\begin{figure}[t]
\centering \epsfxsize=8cm
\epsfbox{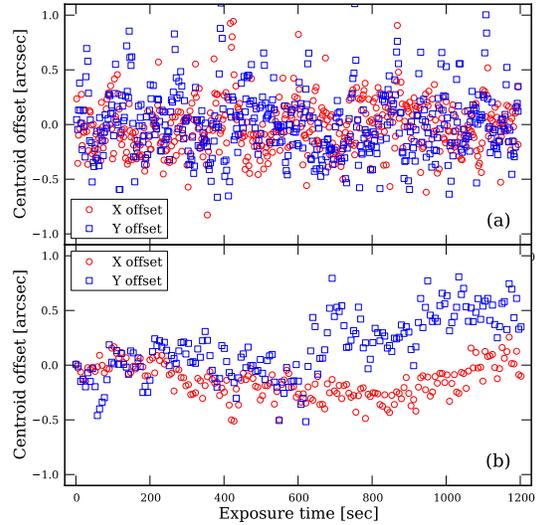}
\caption{Change of centroid position of a guide star on the guide CCD images during 20 minutes.
Panel (a) shows the changes among 1 sec exposures, while panel (b) shows among 5 sec. exposures.
Red open circles denotes the change along X axis of CCD chip, while blue open squares changes along
Y axis.}
\label{guiding}
\end{figure}

We also examined the guiding performance at different guiding rates and exposure times (1 Hz and 0.2 Hz).
Fig. \ref{guiding}
shows the changes of guide star positions on the guide camera during a 1200 sec exposure.  We note guiding corrections made with 1 sec exposure have larger scatter than those from 5 sec exposure. However, auto-guiding with 1 sec exposure was carried out on Aug. 15, a night with worse seeing condition than Aug. 17 when 5 sec exposures were used (See Table 2). Therefore, large scatter may be partly due to the poor seeing condition. The correction for the periodic two minute variation (see Section 2.1) in RA (y-direction here) are clearly seen in the 1 sec guiding, but not in 5 sec guiding. This indicates that guiding with a shorter exposure time is more effective in correcting short-term tracking errors as expected from the tracking performance analysis in Section 2.1. Another factor that needs to be considered is that the tracking errors are known to change with the telescope pointing direction. These considerations make it difficult to assess how much the guiding performance degrades with the increased exposure time. Finally, we report that images taken with 5 sec exposure do not show a serious degradation in the PSF shape, keeping the FWHM about 1$\arcsec$.

\section{CONCLUSION}


We successfully developed, tested, and calibrated an auto-guiding system for CQUEAN. It consists of a camera, an off-axis mirror, and an arm structure to hold all components. To increase the expected number of guide stars that are brighter than the limiting magnitude, we built a mechanism that rotates the guide camera around the optical axis of the telescope.
This is more efficient than adopting a larger format CCD chip fixed to a single position, because the 2.1 m Otto Struve Telescope is a classical cassegrain type which has strong coma aberration. The rotating mechanism minimizes the offset from the optical axis, and thus results in better PSFs in the field.
From the astrometry on the guide CCD camera images, we estimated the offset amount caused by mechanical flexure of the arm and decelerator backlash.
We found 
that these factors do not significantly affect guiding performance. The comparison between two 600 second exposures made on the science camera with and without auto-guiding proves the auto-guider works well and is necessary. Exposures up to 1200 sec long were made with auto-guiding and the resultant images display round stellar profiles.

\acknowledgments
{This work was supported by the National Research Foundation of Korea(NRF) grant funded by the Korean government(MEST) (No. 2009-0063616). E. Kim was partially supported by WCU program through NRF funded by MEST of Korea (R31-10016). The authors thank Changsu Choi, and Juhee Lim for their help in additional test runs of the guide camera, and Yiseul Jeon for her help in data reduction.}

\end{document}